\documentclass[aps]{revtex4}
    \def\be{\begin{equation}}
    \def\ee{\end{equation}}
    \def\ba{\begin{eqnarray}}
    \def\ea{\end{eqnarray}}

    \def\alphat{\tilde{\alpha}}
    \def\betat{\tilde{\beta}}

\begin{document}
\title{Energy-Momentum Tensor of Cosmological Fluctuations during 
           Inflation}

\author{F. Finelli} \email{finelli@bo.iasf.cnr.it}
\affiliation{IASF/CNR, Istituto di Astrofisica Spaziale e 
    Fisica Cosmica \\ Sezione di Bologna \\ Consiglio Nazionale delle 
    Ricerche \\ Via Gobetti 101, I-40129 Bologna -- Italy}

\author{G. Marozzi} \email{marozzi@bo.infn.it}
\author{G. P. Vacca} \email{vacca@bo.infn.it}
\author{G. Venturi} \email{armitage@bo.infn.it}
\affiliation{Dipartimento di Fisica, Universit\`a degli Studi di Bologna
    and I.N.F.N., \\ via Irnerio, 46 -- I-40126 Bologna -- Italy}


\begin{abstract}
We study the renormalized energy-momentum tensor (EMT) of 
cosmological scalar fluctuations during the slow-rollover regime for
chaotic inflation with a quadratic potential and find that it
is characterized by a negative energy density which grows during
slow-rollover. We also approach the back-reaction problem 
as a second-order calculation in perturbation theory finding no 
evidence that the back-reaction of cosmological fluctuations 
is a gauge artifact.  
In agreement with the results for the EMT, the average expansion rate is 
decreased by the back-reaction of cosmological fluctuations. 
\end{abstract}

\maketitle


\section{Introduction}\label{one}
After the latest CMB data \cite{bennett,peiris}, inflation (see 
\cite{lindebook,kolbturner} for a textbook review) seems the 
most promising theory in explaining the large scale 
structure of the Universe. 
According to inflation, the large scale structure of the Universe exhibits 
the fingerprint of quantum fluctuations amplified during the accelerated 
era \cite{all}. On modelling inflation and its transition to the standard 
Big Bang cosmology the constraints on the amplitude and spectrum of 
CMB fluctuation become constraints on the physics of inflation. 
All this modelling depends on the linear treatment of cosmological 
perturbations. Within the framework of inflationary cosmology, on using 
the most recent data, we are 
therefore close to measuring the details of the spectrum of 
fluctuations, while not having a definite idea on the energy content of 
these fluctuations and how they back-react on the inflationary expansion 
responsible for their amplification.

Although cosmological linear perturbation theory during inflation is almost 
a textbook subject, the understanding of nonlinear effects is still at 
the forefront of research. The nonlinearity and gauge invariance of 
General Relativity are tenacious obstacles both at technical and 
interpretational levels, nonetheless the gauge issue has been solved 
to higher orders 
in perturbation theory \cite{bruni,ABML}. Further, many interesting 
effects are appreciable only beyond the linear order. 
From the theoretical point of 
view the back-reaction of gravitational fluctuations on the geometry is one of 
the most interesting issues \cite{tsamis}. Within the inflationary 
context, this problem has 
been tackled by Abramo, Brandenberger and Mukhanov \cite{ABML,ABMP}. The 
intringuing result that the energy-momentum tensor (henceforth EMT) of 
fluctuations may slow down inflation \cite{ABML,ABMP} has subsequently 
generated renewed interest 
in the subject of back-reaction \cite{UNRUH,AW,BRAND}. 
The final answer on the physical significance of this back-reaction 
effect is still under debate \cite{UNRUH,AW,BRAND}. 
The problem of gravitational back-reaction for 
black-holes has also been tackled for gravitational waves \cite{AF}.
More recently these non-linear effects have also drawn attention in 
connection with observations, since non-linear cosmological 
perturbations introduce non gaussian signatures in the power spectrum 
\cite{maldacena,ABMR}.

The aim of this paper is to compute the renormalized EMT of 
cosmological fluctuations during inflation, according to the 
adiabatic regularization scheme \cite{birrell} also used 
in our previous paper \cite{paperI}. The model considered here is the 
slow-rollover regime of inflation driven by a massive inflaton, 
but we believe that the results obtained 
here also hold for other inflationary models. We find that the 
averaged (with respect to the adiabatic vacuum) renormalized 
EMT of cosmological 
fluctuations during slow-rollover is characterized by a negative energy 
density and a de Sitter-like equation of state 
(this result was found for 
long-wavelength modes in \cite{ABML,ABMP}). 
In a naive approach this would lead us 
to think that the EMT of cosmological fluctuations slows down inflation. 
We also evaluate the back-reaction on the geometry in a systematic way by 
proceeding self-consistently to second-order in perturbation theory. In 
order to do this we give a systematic treatment of 
second-order perturbation theory for single scalar field driven inflation 
(see also \cite{ABMR,rigo,nohhwang} for the second-order formalism).
The gauge used for scalar perturbations in this paper is the uniform 
curvature gauge~\cite{hwang} (UCG) generalized to second order. In the 
UCG the spatial sections are not perturbed by scalar fluctuations. 
We believe that this 
gauge is more convenient for our problem than the more frequently used 
longitudinal gauge. 

In this paper we focus on the EMT of scalar cosmological perturbations. 
This is indeed the relevant effect since vector perturbations decay and 
gravitational waves are described by an EMT which is equivalent to a 
massless field \cite{raul}, whose main effect is non-leading with respect to 
scalar fluctuations (see however \cite{tsamis} for
the two-loop calculation). 

The plan of the paper is as follow. In section 2 we present the linear 
cosmological perturbations in the UCG. In section 3 we extend the UCG to 
second order and we give expressions for the Einstein and the 
energy-momentum tensors. In section 4 we present and illustrate the use 
of the Einstein equations to second order. In sections 5 and 7 we 
give the approximate solutions for first and second order fuctuations, 
respectively, using the renormalized values computed in section 6. 
We discuss the back-reaction on the geometry in section 8 and we give our 
conclusions in section 9. In the three appendices we
A) compare our analytical approximation with the WKB method,
B) exhibit the fourth order adiabatic expansion and
C) compare some of our results with those obtained in a different gauge. 


\section{Linear Perturbations in the Uniform Curvature Gauge} \label{two}
We consider inflation in a flat universe driven by a classical minimally 
coupled scalar field with a general potential $V(\phi)$. 
The action is:
    \be
    S \equiv \int d^4x {\cal L} = \int d^4x \sqrt{-g} \left[ 
\frac{R}{16{\pi}G}
    - \frac{1}{2} g^{\mu \nu}
    \partial_{\mu} \phi \partial_{\nu} \phi
    - V(\phi) \right] \,
    \label{action}
\ee
where ${\cal L}$ is the Lagrangian density.

Let us now study the fluctuations of the scalar field $\varphi (t,{\bf 
x})$ around its homogeneous classical
\footnote{For a quantum treatment 
of the homogeneous inflaton see \cite{chaotic,time}.} value $\phi (t)$ and 
include metric perturbations. For the homogeneous case we have 
\begin{eqnarray}
&& \ddot{\phi}+3H\dot{\phi}+V_\phi=0 \nonumber \\
&& H^2 = \frac{8 \pi G}{3} \left[ \frac{\dot \phi^2}{2} + V \right]
\end{eqnarray}
where $H=\dot a /a$ is the Hubble parameter and $a$ is the scale factor.

The scalar perturbations around a flat Robertson-Walker metric are:
\be
ds^2=-(1+2 \alpha)dt^2-a \beta_{,i} dt dx^i+a^2 \left [ 
\delta_{ij}(1 - 2 \psi)+2 \gamma_{,ij} \right ]dx^i dx^j \,,
\label{PER_GEN2}
\ee 
where the symbol $,_i$ denotes the derivative 
with respect to the spatial coordinates.
We choose to work in the uniform curvature gauge:
\be
ds^2=-(1+2 \alpha)dt^2 - a \beta_{,i} dt dx^i+a^2 \delta_{ij} dx^i dx^j 
\,.
\label{PER_UCG2}
\ee
We note that this choice fixes uniquely the gauge, just as the more 
frequently used longitudinal gauge (for a review of cosmological 
perturbations in this gauge see \cite{MFB}). This can be seen 
by setting $\epsilon^0 = - \psi/H$, $\epsilon = \gamma$, 
where $\epsilon^\mu = (\epsilon^0, \epsilon^{,i})$ is an infinitesimal 
coordinate transformation ($x^\mu \rightarrow x^\mu + \epsilon^\mu$). 
In order to see the connection with the more known longitudinal gauge we 
write the metric \cite{MFB} in that gauge:
\be
ds^2=-(1 + 2 \Phi)dt^2 + a^2 (1 - 2 \Psi) \delta_{ij} dx^i dx^j
\label{longitudinal}
\ee
and note that the transformation between the two gauges can be obtained 
through a time reparametrization $\epsilon^0 = - a \beta/2$, $\epsilon=0$:
\be
\Phi = - \alpha + \frac{d}{dt} \left( \frac{a}{2} \beta \right)
\,, \quad \Psi = - \frac{a H}{2} \beta \,.
\ee

Let us now derive the equation of motion in the uniform curvature gauge 
(see also \cite{hwang}~\footnote{We observe that the equations in 
\cite{hwang} are for a metric perturbation which has $g_{0i} = a 
\beta,_i$ and not $g_{0i} = a \beta,_i/2$ as is stated there.}). 
The scalar field fluctuations obey to the 
following equation of motion:
\begin{eqnarray}
\ddot{\varphi}+3 H \dot{\varphi}-\frac{1}{a^2}\nabla^2 \varphi + 
V_{\phi \phi}
\varphi &=& 2 \alpha \ddot \phi + \dot \phi \left( \dot \alpha + 6 H 
\alpha - \frac{1}{2a} \nabla^2\beta \right)
\nonumber \\
&=& \dot{\alpha}\dot{\phi} - 2 \alpha V_\phi 
-\frac{\dot{\phi}}{2a} \nabla^2\beta \,, 
\label{EQMP4}
\end{eqnarray}
where a dot denotes a derivative w.r.t. the time $t$.
Starting from the Einstein equations, $G^{\mu}_{\nu}=8\pi G
T^{\mu}_{\nu}$,
in order to obtain an equation for 
$\varphi$ only one needs the energy and momentum constraints in their 
linearized version, i.e. the $G^0_0$ and $G^0_i$ linear equations:
\begin{eqnarray}
\frac{H}{a} \nabla^2 \beta &=& 8 \pi G \left ( \dot{\phi}
\dot{\varphi} + V_\phi \varphi + 2 \, V \alpha \right)  
\nonumber \\
&=& 8 \pi G \frac{\dot{\phi}^2}{H} \, \frac{d}{dt} \left
(\frac{H}{\dot{\phi}}\varphi \right )\,,
\label{Eq_beta} \\
\alpha_{,i} &=& 4 \pi G \frac{\dot{\phi}}{H} \, \varphi_{,i} 
\label{Eq_alpha} 
\end{eqnarray}
Because of the absence of anisotropic stress we also have $\dot \beta + 2 
H \beta = 2 \alpha / a$ (this relation replaces the equality 
$\Phi = \Psi$ in the longitudinal gauge) \cite{hwang}. On subtituting 
these two latter equations in Eq. (\ref{EQMP4}) one obtains 
\be
\ddot{\varphi} + 3 H \dot{\varphi} - \frac{1}{a^2}\nabla^2 \varphi +
\left[ V_{\phi \phi} + 2 \frac{\dot H}{H}\left(3 H - 
\frac{\dot H}{H} + 
2 \frac{\ddot{\phi}}{\dot{\phi}}\right )\right] \varphi = 0 \,.
\label{Eq_mukhanov}
\ee
It is important to note that the effective potential for the fluctuations 
can be rewritten as:
\be
V_{\phi \phi} + 2 \frac{\dot H}{H} \left(3 H -
\frac{\dot H}{H} +
2 \frac{\ddot{\phi}}{\dot{\phi}} \right) = V_{\phi \phi} - 6 H^2 \left( 
\epsilon -\frac{1}{3}\epsilon^2 + \frac{\dot \epsilon}{3 H} \right) \,,
\label{gravbr}
\ee
where we have introduced the (positive) slow-rollover parameter $\epsilon 
\equiv 
- \dot H/H^2$.
This means that the self-consistent inclusion of gravitational 
fluctuations changes the effective potential 
for the field fluctuations. In particular, gravitational fluctuations 
generally decrease the effective mass to first order in $\epsilon$. 

Equation (\ref{Eq_mukhanov}) is generally seen written in conformal 
time. The Fourier tranform modes of $v = a \varphi$ satisfy:
\be
v_k'' + \left( k^2 - \frac{z''}{z} \right) v_k = 0 \,, \quad z = a 
\frac{\dot \phi}{H} \,,
\label{Eq_resc_mukhanov}
\ee
where $'$ denotes a derivative with respect to the conformal time $\eta$, 
$d \eta = d t/a$. 
On comparing the last equation with Eq. (12) of \cite{mukhanov}
it is immediate to see 
that $\varphi$ satisfies the 
same equation as the Mukhanov variable $Q$. Therefore, 
the uniform curvature gauge has the advantage of singling out the 
true dynamical degrees of freedom (the matter ones), even if it has the 
disadvantage of being non diagonal in the metric perturbations.  


\section{Beyond the Linear Order} \label{three}
To second order we consider a metric having the following coefficients:
\begin{eqnarray}
g_{00} &=& - 1 - 2 \alpha - 2 \alpha^{(2)} \nonumber \\
g_{0i} &=& - \frac{a}{2} \left( \beta_{,i} + \beta_{,i}^{(2)} \right) 
\nonumber \\
g_{ij} &=& a^2 \left[ \delta_{ij} + \frac{1}{2} \left( \partial_i 
\chi_{j}^{(2)} + 
\partial_j \chi_{i}^{(2)} + h_{ij}^{(2)} \right) \right] \,.
\label{metric_second}
\end{eqnarray}
The above metric is the extension of the uniform curvature gauge to second 
order: $\alpha^{(2)}$ and $\beta^{(2)}$ are the scalar perturbations to  
second order. To second order, scalar, vector and tensor perturbations do 
not evolve independently as is the case in first order. For this reason we 
take into account second order vector and tensor perturbations, 
represented by 
the divergenceless vector $\chi_j^{(2)}$ and by the transverse and 
traceless tensor $h_{ij}^{(2)}$, respectively. In the above we have 
omitted 
first vector perturbations (which die away kinematically) and tensor 
perturbations (which satisfy the usual equation $\ddot h + 3 H \dot h - 
\nabla^2 h /a^2 = 0$). With this approximation we are neglecting the EMT of 
vector and tensor perturbations, and their correlations with the scalar 
perturbations.
We finally note that the choice in Eq. (\ref{metric_second}) (including 
vector and tensor metric elements to first order) fixes the 
gauge completely to second order.

The Einstein tensor expanded to second order is:
\begin{eqnarray}
G^0_0 &=& 
G^{0 (0)}_{0} + \delta G^{0 (1)}_{0} + \delta
G^{0 (2)}_{0} \nonumber \\
&=&
- 3 H^2 - \frac{H}{a} \nabla^2 \beta + 6 H^2 \alpha
\nonumber \\
& & - \frac{H}{a} \nabla^2 \beta^{(2)} + 6 H^2 \alpha^{(2)} - 12
H^2 \alpha^2 + \frac{3}{4} H^2 |\vec{\nabla} \beta|^2
\nonumber \\
& & +\frac{H}{a} \left( \vec{\nabla}\alpha
\cdot \vec{\nabla} \beta + 2 \alpha \nabla^2 \beta \right) \nonumber \\
& & + \frac{1}{8 a^2} \left( \beta_{,ij} \beta^{,ij} - \left(\nabla^2 
\beta \right)^2 \right ) 
\label{GE_00} \\
G^0_{i} &=& G^{0 (0)}_{i} + \delta G^{0 (1)}_{i} + \delta 
G^{0 (2)}_{i} \nonumber \\
&=& - 2 H\alpha_{,i} - 2 H \alpha_{,i}^{(2)} + 8 H \alpha \alpha_{,i}
- \frac{1}{2 a} H \alpha_{,i}\nabla^2 \beta
\nonumber \\
& & + \frac{1}{2 a} \vec{\nabla} \alpha \cdot \vec{\nabla} \beta_{,i}
- \frac{H}{2} \vec{\nabla} \beta_{,i} \cdot \vec{\nabla} \beta - 
\frac{1}{4} \nabla^2 \dot \chi_i^{(2)} 
\label{GE_0i} \\
G^i_{j} &=& G^{i (0)}_{j} + \delta G^{i (1)}_{j} + \delta G^{i (2)}_{j}
\nonumber \\
&=& \delta^i_j \left\{ - (3H^2+2\dot{H}) 
+ 2 \alpha (3H^2+2\dot{H})+ 2 H \dot{\alpha} \right.
\nonumber \\ 
& & \left. + \frac{1}{a^2} \nabla^2 \alpha
- \frac{H}{a} \nabla^2 \beta
-\frac{1}{2 a} \nabla^2 \dot{\beta} \right.
\nonumber \\
& & \left. + 2 \alpha^{(2)} (3H^2+2\dot{H})+ 2 H \dot{\alpha}^{(2)}
+ \frac{1}{a^2} \nabla^2 \alpha^{(2)} 
- \frac{H}{a} \nabla^2 \beta^{(2)}
-\frac{1}{2 a} \nabla^2 \dot{\beta}^{(2)}
\right. \nonumber \\
& & \left. + \frac{H}{a} \vec{\nabla}\alpha \cdot \vec{\nabla} \beta 
+\frac{H}{2} \vec{\nabla}\beta \cdot \vec{\nabla} 
\dot{\beta}+\left (\frac{1}{4}
|\vec{\nabla} \beta|^2-4\alpha^2 \right ) \left(3H^2+2\dot{H}\right )- 
\right.
\nonumber \\
& & \left. - 8 H \alpha\dot{\alpha} + \left
(\frac{\dot{\alpha}}{2 a} +2 \frac{H}{a} \alpha\right ) 
\nabla^2 \beta
- \frac{2}{a^2} \alpha \nabla^2 \alpha + \frac{\alpha}{a}
\nabla^2 \dot{\beta} - \frac{1}{a^2} |\vec{\nabla} \alpha|^2 
\right.
\nonumber \\
& & \left.+ 
\frac{1}{8 a^2} \left( \beta^{, \ell m} \beta_{, \ell m} - ( \nabla^2 
\beta )^2 \right) \right\} + \left\{ \frac{1}{2 a} \dot \beta^{, i}_{, j} 
+ \frac{H}{a} \beta^{, i}_{, j} - \frac{1}{a^2} \alpha^{, i}_{, j} \right.
\nonumber \\ & & 
\left. +\frac{1}{2 a} \dot \beta^{(2) \, , i}_{\,\,\,\,\,\,\,\,\,, j}
+ \frac{H}{a} \beta^{(2) \,, i}_{\,\,\,\,\,\,\,\,\,, j} -
\frac{1}{a^2} \alpha^{(2) \,, i}_{\,\,\,\,\,\,\,\,\,, j}
+ \frac{1}{a^2} \alpha^{, i} \alpha_{, j} - \frac{H}{a} \beta^{, i}
\alpha_{, j} + \frac{2}{a^2} \alpha \alpha^{, i}_{, j}
\right.
\nonumber \\ & &
\left.
- \frac{2}{a} H 
\alpha \beta^{, i}_{, j} - \frac{1}{2 a} 
\dot \alpha \beta^{, i}_{, j} - \frac{1}{a}
\alpha \dot{\beta}^{, i}_{, j} + \frac{1}{4 a^2} \left( \nabla^2 
\beta \beta^{, i}_{, j} - \beta^{,i}_{, k } \beta^{,k}_{, j}
\right) \right.
\nonumber \\ & & \left. + \frac{3}{4} H \left (
\dot{\chi}^{(2),i}_{j}+\dot{\chi}^{(2) i}_{\,\,\,\,\,\,\,\, ,j}
+\dot{h}^{(2) i}_{\,\,\,\,\,\,\, j} \right )+ \frac{1}{4}
\left(\ddot{\chi}_{j}^{(2),i}+\ddot{\chi}^{(2) i}_{\,\,\,\,\,\,\,\,
  ,j}+\ddot{h}^{(2) i}_{\,\,\,\,\,\,\, j} \right) \right.
\nonumber \\& & \left. - \frac{1}{4 a^2} \nabla^2 h^{(2) i}_{\,\,\,\,\,\,\, j}
\right\}
\label{GE_ii} 
\end{eqnarray}

To second order the scalar field is expanded as:
\be
\phi (t, {\bf x} ) = \phi (t) + \varphi (t, {\bf x} ) + \varphi^{(2)} (t,
{\bf x} )
\label{second_field}
\ee
The EMT of inflaton fluctuations 
\be
T^\mu_\nu = \partial^\mu \phi \partial_\nu \phi + \delta^\mu_\nu {\cal L}
\ee
to second order is:

\begin{eqnarray}
T_{0}^{0} &=& T^{0 \, (0)}_0+\delta T^{0\,(1)}_0+\delta T_{0}^{0 \, 
(2)}
\nonumber \\
&=& - \left[ \frac{1}{2} 
\dot{\phi}^2 + V(\phi) \right] + \dot \phi^2  \alpha - \dot{\phi}
\dot{\varphi}
- V_\phi \varphi
\nonumber \\
& & + \dot \phi^2 \alpha^{(2)}- \dot{\phi}
\dot{\varphi^{(2)}}
- V_\phi \varphi^{(2)} - \frac{1}{2} \dot{\varphi}^2
- \frac{1}{2a^2}|\vec{\nabla} \varphi|^2
- \frac{1}{2} V_{\phi \phi} \varphi^2 \nonumber \\
& &- 2 \dot \phi^2 \alpha^2 +
2 \dot \phi \alpha \dot \varphi + \frac{1}{8}\dot{\phi}^2
|\vec{\nabla} \beta|^2
\label{TEIP_00} \\ & & \nonumber \\
T^0_{i} &=& T^{0\,(0)}_i +\delta T^{0\,(1)}_i+\delta T^{0\,(2)}_i
\nonumber \\
&=& - \dot{\phi} \varphi_{,i} - \dot{\phi} \varphi^{(2)}_{,i} - 
\dot{\varphi} \varphi_{,i} + 2 \dot{\phi} \alpha \varphi_{,i}
\label{TEIP_0i} \\
T^i_{j} &=& T^{i \, (0)}_j +\delta 
T_{ii}^{(1)}+\delta T_{ii}^{(2)} 
\nonumber \\
&=& \left[ \frac{1}{2} \dot{\phi}^2 - V \right] \delta^i_j 
+ \left[ \dot{\phi}\dot{\varphi} - \dot{\phi}^2  \alpha 
- V_\phi \varphi \right] \delta^i_j
+ \left[ \dot{\phi} \dot{\varphi}^{(2)} - \dot{\phi}^2 \alpha^{(2)}
- V_\phi \varphi^{(2)} 
\right.
\nonumber \\
& & \left.
+ \frac{1}{2} \left( \dot{\varphi}^2
- 4\dot{\phi}\alpha\dot{\varphi} - \frac{1}{4}\dot{\phi}^2|\vec{\nabla}
\beta|^2 + 4\dot{\phi}^2\alpha^2
+ \frac{1}{a}\dot{\phi}\vec{\nabla}\varphi \cdot \vec{\nabla}
\beta - \frac{1}{a^2}|\vec{\nabla} \varphi|^2
- V,_{\phi \phi} \varphi^2
\right) \right] \delta^i_j \nonumber \\
& & + \frac{1}{a^2} \varphi^{,i} \varphi_{,j} - \frac{\dot \phi}{2 a} 
\beta^{,i} \varphi_{,j}
\label{TEIP_ii}
\end{eqnarray}


\section{Equations to Second Order} \label{four}
In this section we exhibit the second order equations corresponding 
to Eqs. (\ref{Eq_beta},\ref{Eq_alpha},\ref{Eq_mukhanov}), using, when
convenient, the homogeneous and first order equations.
We first give the momentum constraint to second order, which is
obtained from the expressions given in the Eqs. (\ref{GE_0i})
and (\ref{TEIP_0i}):
\be
\alpha^{(2)}_{,i} + \frac{1}{8 H} \nabla^2 \dot \chi_i^2 = 4 \pi G
\frac{\dot \phi}{H} \varphi^{(2)}_{,i} + S_i
\ee
where $\chi_i$ are vector metric perturbations and:
\begin{eqnarray}
S_i =
\frac{4\pi G}{H} \left( \dot{\varphi} - 2\dot{\phi} \, \alpha \right)
\varphi_{,i} +4 \alpha \alpha_{,i} -
\frac{1}{4aH} \left( \alpha_{,i} \nabla^2 \beta-
\alpha^{,j} \, \beta_{,ij} \right)-
\frac{1}{4} \beta^{,j} \, \beta_{,ij}
\label{Si}
\end{eqnarray}
The term $S_i$ can be written as $\partial_i s + v_i$, further
\be
\alpha^{(2)} = 4 \pi G
\frac{\dot \phi}{H} \varphi^{(2)} + s 
\label{alpha2}
\ee
where
\begin{eqnarray}
s &\equiv& - 4 \pi G \epsilon \varphi^2 + 2 \alpha^2 - \frac{1}{8} | 
\vec{\nabla} \beta|^2 
\nonumber \\
&&+ \frac{1}{\nabla^2} \left[\frac{ 4 \pi G }{H} \vec{\nabla}
\cdot \left( \dot \varphi \vec{\nabla} \varphi \right) + \frac{1}{4 a H} \left(
\alpha^{,kj} \beta_{,kj} - \alpha^{,k}_{,k} \beta^{,j}_{,j} \right)
\right]
\label{s_definition}
\end{eqnarray}
contains the quadratic contribution of first order perturbations. We note 
that $s$ includes non-local spatial contributions which nevertheless
to leading order in $\epsilon$ and for long wavelenght
have an ordinary behaviour on large scales. One can in fact approximate
in such a limit, for the
isotropic case, the first term in the second line of eq.
(\ref{s_definition}) as $(2\pi G)/H \, \varphi \dot{\varphi}$. 
On combining Eqs. (\ref{GE_00},\ref{TEIP_00}),
the energy constraint to second order, upon using Eq. (\ref{alpha2})
and the lower order constraints, is given by:
\be
\frac{H}{a} \nabla^2 \beta^{(2)} 
= 8 \pi G \frac{\dot{\phi}^2}{H} \, \frac{d}{dt} \left
(\frac{H}{\dot{\phi}} \varphi^{(2)} \right) - Q + 16 \pi G V s
\,,
\label{Eq_beta_second} 
\ee
where $Q$ is defined as:
\begin{eqnarray}
Q &=& 12 H^2 \alpha^2 -\frac{3}{4} H^2 |\vec{\nabla} \beta|^2
-\frac{H}{a} \Bigl( \alpha_{,i} \, \beta^{,i} +
2\alpha  \nabla^2 \beta \Bigr) -
\frac{1}{8 a^2} \Bigl( \beta_{,ij} \, \beta^{,ij}
-\left( \nabla^2 \beta \right)^2\Bigr) 
\nonumber \\
&&- 8 \pi G \Biggl[ \frac{1}{2}\dot{\varphi}^2 +\frac{1}{2 a^2}
|\vec{\nabla} \varphi|^2 +\frac{1}{2} V_{\phi\phi} \varphi^2 +
2 \dot{\phi}^2 \alpha^2 -2\dot{\phi} \alpha \dot{\varphi}
-\frac{1}{8} \dot{\phi}^2 |\vec{\nabla} \beta|^2 \Biggr]
\label{Q}
\end{eqnarray}
We observe that Eq. (\ref{Eq_beta_second}) for $\beta^{(2)}$ is 
reminiscent of a universe filled with two components, with the term
$Q-16 \pi G V s$ playing the role of a non-adiabatic pressure term~\cite{MFB}, 
upon comparing with the first order equation (\ref{Eq_beta}). 
From this analogy one may guess that the terms on the right hand side of 
Eq. 
(\ref{Eq_beta_second}) should approximatively cancel.
Explicit calculations confirm this property.

The equation of motion for the scalar field to second order, after
using all the previous constraints, is given by:
\be
\ddot{\varphi}^{(2)} + 3 H \dot{\varphi}^{(2)} - \frac{1}{a^2}\nabla^2 
\varphi^{(2)} + \left[ V_{\phi \phi} + 2 \frac{\dot H}{H}\left(3 H -
\frac{\dot H}{H} +
2 \frac{\ddot{\phi}}{\dot{\phi}}\right )\right] \varphi^{(2)} = D \,,
\label{Eq_mukhanov_second}
\ee
with an homogeneous part which is the same as found, see
Eq. (\ref{Eq_mukhanov}), for the first order fluctuations and an
inhomogeneous term $D$ which is purely quadratic in terms of the first
order fluctuations. In particular one obtains
\be
D = R + \dot \phi \dot s - 2 V_{\phi} s +  \frac{\dot \phi}{2 H} 
\left( Q - 16 \pi G V s \right)
\ee
 and 
\begin{eqnarray}
R=&-& \,\frac{1}{2} V_{\phi\phi\phi} \varphi^2 -2\dot{\phi}\, \alpha
\dot{\alpha} + \dot{\alpha} \dot{\varphi} +
\frac{2}{a^2} \alpha \nabla^2 \varphi -2 V_{\phi\phi} \alpha \varphi +
\frac{\dot{\phi}}{2a} \alpha_{,i} \, \beta^{,i} -
\frac{1}{4} V_{\phi}|\vec{\nabla} \beta|^2+\nonumber \\
&+&\frac{1}{a^2} \alpha_{,i} \, \varphi^{,i}-
\frac{H}{a} \beta_{,i} \, \varphi^{,i}-
\frac{1}{2a} \dot{\varphi} \nabla^2 \beta+
\frac{\dot{\phi}}{4} \beta_{,i} \, \dot{\beta}^{,i}-
\frac{1}{2a}  \varphi_{,i} \, \dot{\beta}^{,i}-
\frac{1}{a} \beta_{,i} \, \dot{\varphi}^{,i}
\label{R}
\end{eqnarray}

We have therefore obtained the perturbative equation for the scalar field 
fluctuations to second order, which is a novel result.


\section{Approximate Solution for Linear Perturbation} \label{five} 
The equation for the Mukhanov variable (\ref{Eq_resc_mukhanov}) does not 
have an exact solution, except for the known case of an exponential 
potential 
\cite{exponential}. Approximate schemes to obtain the long wavelength 
solution, such as the slow-rollover technique \cite{SL}, exist.
Any approximation must agree with the solution to Eq. 
(\ref{Eq_resc_mukhanov}) with $k=0$:
\be
v_{| k=0} = C z + D z \int \frac{d t}{a z^2}
\label{infinite}
\ee
where $C$ and $D$ are constants (when the first term in Eq. 
(\ref{infinite}) dominates, the curvature perturbation $\zeta = H \varphi 
/ \dot \phi$ is constant to leading order). For small wavelengths ($k \sim 
0$) $C$ and $D$ include the $k$ dependence of the modes.
However, in order to perform an adiabatic 
regularization, we do not only need the solution for small $k$, but 
for the whole spectrum. The slow rollover approximation applied to Eq. 
(\ref{Eq_resc_mukhanov}) leads to a Bessel function whose infrared limit 
agrees with Eq. (\ref{infinite}) only on freezing a spurious time 
dependence of the Bessel solution to the value on horizon crossing. This 
problem is the same as was encountered in \cite{paperI} on trying to extend 
the approximate solution for the inflaton fluctuations in a rigid 
space-time to fields with different mass. Therefore, the slow rollover 
paradigm is not useful in order to obtain an approximate solution over the 
whole $k$ spectrum.

In order to consider the slow-rollover of massive chaotic inflation
we take $\dot H \simeq - m^2/3$, $\ddot H \simeq 0$ and neglect terms 
${\cal O} (\dot H^2/H^4)$ \cite{paperI}. During slow-rollover the equation 
for $\varphi$ is then:
\be
\ddot{\varphi}_k+3 H \dot{\varphi}_k+
\left [\frac{k^2}{a^2}+m^2+6 \dot{H} \right ] \varphi_k=0 \,.
\label{EQMPF_k}
\ee
On comparing with the fluctuations in a rigid spacetime considered in 
\cite{paperI}, we see that the gauge-invariant fluctuations in Eq. 
(\ref{Eq_mukhanov}) 
have a negative mass since $\dot H \simeq -m^2/3$: this is due to the proper 
inclusion of gravitational fluctuations. This result is true for many 
inflationary models and it is related to the red spectrum of the curvature 
perturbations.

In order to have an approximate solution for the whole $k$ spectrum we 
proceed in analogy with \cite{paperI} and we choose for large $k$:
\be
\varphi_k = \frac{1}{a^{3/2}} \left ( \frac{\pi \lambda}{4H}\right )^{1/2}  
H_\nu^{(1)} (\lambda \xi)
\label{ultraviolet}
\ee
where $\xi = k/(a H)$ and 
\begin{eqnarray}
\lambda &=& 1 - \frac{\dot H}{H^2} = 1 + \epsilon \nonumber \\  
\nu &=& \frac{3}{2} - \frac{1}{3} \frac{m^2}{H^2} - 3 \frac{\dot H}{H^2} = 
\frac{3}{2} + \frac{2}{3} \frac{m^2}{H^2} = \frac{3}{2} + 2 \epsilon \,.
\end{eqnarray}
We note that $\nu$ for the gauge-invariant 
fluctuations differs from the corresponding quantity for inflaton 
fluctuations in rigid space-times \cite{paperI}.

The reason the procedure followed in \cite{paperI} does not succeed in 
producing an approximate solution, valid over the whole range of $k$, is the 
dependence of $\nu$ on $H$. In fact, among the terms which are 
apparently ${\cal O} (\dot 
H^2/H^4)$, the following term:
\be
\sim {\dot H}{H^2} \frac{\partial \nu}{\partial H} \ln \zeta
\ee
leads to a term which is of order ${\cal O} (\dot H/H^2)$
\be
\sim{\cal O} (\frac{\dot H^2}{H^4})
\left [\ln \frac{k}{H} +\frac{1}{2}
\left (\frac{H_0^2-H^2}{\dot{H}}\right)\right] \,,
\ee
making the approximation not valid for small $k$ and at large times.
For this reason we consider Eq. (\ref{ultraviolet}) only as an 
approximation for modes which are still inside the Hubble radius.
We also note that the solution for the ultraviolet regime in 
Eq. (\ref{ultraviolet}) is the same as that of the slow-rollover 
approximation, although here $\dot H \simeq {\rm const.}$ and not
$\epsilon \simeq {\rm const.}$

On considering the infrared limit for the gauge-invariant fluctuations 
in Eq. (\ref{infinite}) we know that $\varphi_k \sim {\cal O} (1/H)$, 
since $\dot \phi \sim {\rm const.}$ For $k \le a H$ we must to replace the 
solution in Eq. (\ref{ultraviolet}) in order to reproduce the correct 
behaviour in the infrared by:
\be
\varphi_{\bf k} = \frac{1}{a^{3/2}}
\left(\frac{\pi \lambda}{4H}\right )^{1/2} \left( \frac{H(t_k)}{H(t)}
\right )^2 H_{3/2}^{(1)} (\lambda \xi) \,, 
\label{solution_infrarossa_in} 
\ee
where $t_k$ is a time related to the istant for which the mode $k$ crosses 
the Hubble radius. We leave the details of estimating the time $t_k$ to 
appendix A and give the Hubble parameter for this time value
\be
H(t_k)= H_0 \sqrt{1+2\frac{\dot{H}_0}{H_0^2} \log{\frac{\lambda_0 k}{H_0 
\nu_0}}} \,.
\ee
We note that we have a nearly scale invariant spectrum
due to the dependence on $t_k$.

We also note that the amplitude of the curvature 
perturbations, $\zeta_k = H \varphi_k/ \dot \phi$, associated with the 
solution in Eq. (\ref{solution_infrarossa_in}), need not obey the 
observational constraints ($k^{3/2} |\zeta_k| \sim 10^{-5}$). 
The reason is that the spectrum is red tilted and we 
are considering modes which could have exited the Hubble radius much 
earlier than the ones relevant for observations. 
Only if the duration of inflation is minimal in order to solve the horizon 
problem do the curvature perturbations associated with Eq. 
(\ref{solution_infrarossa_in}) satisfy the observational constraints. 

Cosmological fluctuations are canonically quantized as usual and we shall 
consider the vacuum defined by Eqs. 
(\ref{ultraviolet}) and (\ref{solution_infrarossa_in}): 
\be
\hat{\varphi} (t, {\bf x}) = \frac{1}{(2 \pi)^3} \sum_{\bf k}
\left[ \varphi_{k} (t) \, e^{i {\bf k} \cdot {\bf x}} \,\, \hat{b}_{\bf k} + 
 \varphi^*_{k} (t) e^{- i {\bf k} \cdot {\bf x}} \,\,
\hat{b}^\dagger_{{\bf k}} \right]
\label{quantumFourier}
\ee
where the $\hat{b}_k$ are time-independent Heisenberg operators.
 

\section{The Adiabatic Subtraction} \label{six}
We now compute the integrals obtained on taking the expectation values of 
the relevant operators. The strategy is slightly 
different from that of our previous calculation \cite{paperI} since the mode 
function is obtained by matching two approximate solutions. Therefore we 
split the integral:
\be
\int_{\ell}^{+\infty} dk\rightarrow \int_{\ell}^{\nu 
\frac{aH}{\lambda}} dk
+\int_{\nu \frac{aH}{\lambda}}^{+\infty} dk
\ee
where $\ell=C H_0$ is an infrared cut-off related to the beginning of 
inflation~\footnote{see also \cite{paperI} for numerical considerations
on the initial states.} \cite{VF,cutoff} and $k=\nu \frac{aH}{\lambda}$ 
is the turning-point of the Bessel function in Eq.(\ref{ultraviolet}).
We substitute the infrared solution 
(\ref{solution_infrarossa_in}) in the first integral and the ultraviolet 
solution (\ref{ultraviolet}) in the second integral.
For the ultraviolet part we employ, as in the previous article  \cite{paperI},
dimensional regularization using a $d$ dimensional space measure. 
Subsequently an adiabatic subtraction is performed in order to obtain the
renormalized quantities. 

From our previous calculation we know that the leading contributions come 
from terms which contain $\langle \varphi^2 \rangle$, while terms such as  
$\langle \nabla_\mu \varphi \nabla^\mu \varphi \rangle$ are more 
ultraviolet and therefore non-leading. By the symbol $\langle ... 
\rangle$ we denote the average over the quantum state defined by Eqs.
(\ref{ultraviolet}) and (\ref{solution_infrarossa_in}). 
It is important to note that the 
standard fourth order adiabatic subtraction is sufficient to regularize 
the EMT of cosmological fluctuations as well. In fact, the terms which 
apparently would not be regularized by a fourth order expansion (see 
the term $\beta_{,ij} \, \beta^{,ij}
-\left( \nabla^2 \beta \right)^2$ in the $G^0_0$ equation 
(\ref{GE_00}), for instance) vanish on averaging over an homogeneous state.   
As an example of the calculations we exhibit the details of the calculation 
of $\langle\varphi^2\rangle_{\rm REN}$. 

The ultraviolet and infrared integrals are respectively:
\begin{eqnarray}
\langle\varphi^2\rangle^{UV}&=&\frac{\hbar}{16\pi^2} H^2
\left \{ A+B \frac{m^2}{H^2}+C \frac{H^2}{m^2} -\left ( 2+\frac{2}{3} \frac{m^2}{H^2}\right )
\left( \frac{1}{2 \pi^{1/2}} \right)^{d-3} \left(\nu \frac{a H}{\lambda}\right)^{d-3} \Gamma
\left (\frac{1}{2}-\frac{d}{2}\right )\right\}+ \nonumber \\
& & +{\cal O} (d-3)\,,
\label{phi2_UV}
\end{eqnarray}
\begin{eqnarray}
\langle\varphi^2\rangle^{IR}&=&\frac{\hbar}{4\pi^2}
\frac{1}{\lambda^2} \left (\frac{H_0}{H(t)}\right )^4 H^2
\frac{1}{6 \epsilon_0}
\left\{ \left( 1-2\epsilon_0 \log \frac{\lambda l}{H\nu}\right)^3
-\left(1-2\epsilon_0 \log a\right)^3\right\} \, ,
\label{phi2_IR}
\end{eqnarray} 
where $\epsilon_0=-\dot{H}/H_0^2$ and
$A$, $B$ and $C$ are constants with a complicated dependence on hypergeometric functions 
and we have taken for the calculation of Eq.(\ref{phi2_IR})
$\frac{\lambda_0}{H_0 \nu_0}\simeq \frac{\lambda}{H \nu}$ as is done for the case of the 
$t_k$ derivation in appendix A.
At the end of inflation $\log a \to 1/(2\epsilon_0)$ which is the
inverse of the slow-rollover perturbation parameter. Therefore in the
above expression, in order to be as accurate as possible towards
 the end of inflation
we have included all the leading $\log$ terms which are multiplied
by the small parameter $\epsilon_0$. 
The adiabatic fourth order is (see appendix B and \cite{paperI} ):
\begin{eqnarray}
\langle\varphi^2\rangle_{(4)}&=&\frac{\hbar}{16\pi^2}
H^2 \left \{ \frac{127}{45}-\frac{43}{90}\frac{m^2}{H^2}+
\frac{29}{15}\frac{H^2}{m^2}-
\left (2+\frac{2}{3}\frac{m^2}{H^2}\right )
\left( \frac{1}{2 \pi^{1/2}} \right)^{d-3} \right. \nonumber \\
& & \left. \,(a m)^{d-3} \Gamma
\left (\frac{1}{2}-\frac{d}{2}\right) + {\cal O}\left(\frac{1}{a^3} \right
) \right\} +{\cal O} (d-3)\,,
\label{phi2_adiabatico_5}
\end{eqnarray}
and the resulting renormalized quantity is 
\begin{eqnarray}
\langle\varphi^2\rangle_{\rm REN} &=&\lim_{d\rightarrow 3}
\left(\langle\varphi^2\rangle^{IR}+\langle\varphi^2\rangle^{UV}
-\langle\varphi^2\rangle_{(4)} \right ) \nonumber \\
&=&\frac{\hbar}{4\pi^2}
\frac{1}{\lambda^2} \left (\frac{H_0}{H(t)}\right )^4 H^2
\frac{1}{6\epsilon_0}
\left\{ \left( 1-2\epsilon_0 \log \frac{\lambda l}{H\nu}\right)^3
-\left(1-2\epsilon_0 \log a\right)^3 \right\}
+  \nonumber \\
& & +\frac{\hbar}{16\pi^2} H^2
\left \{ A+B \frac{m^2}{H^2} +C \frac{H^2}{m^2}- 
4 \left (1+\frac{1}{3}\frac{m^2}{H^2}\right )
 \ln \left(\frac{\nu H}{\lambda m}\right)
-\frac{127}{45}+\frac{43}{90}\frac{m^2}{H^2}-\frac{29}{15}\frac{H^2}{m^2} + 
\right. \nonumber \\
& & \left. + {\cal O}\left(\frac{1}{a^3}\right )
\right\}\,.
\label{phi2_rinormalizzato_cap5}
\end{eqnarray} 

The leading behaviour of the renormalized correlator at the beginning
of inflation, 
which actually originates in the infrared region, grows with $\log a$:
\be
\langle\varphi^2\rangle_{\rm REN} \sim 
\frac{\hbar}{4 \pi^2} H^2 \left (\frac{H_0}{H}\right )^4 
\ln a  \,.
\label{phi2_rinormalizzato_leading}
\ee
Its time derivative is 
\be
\frac{d}{d t} \langle\varphi^2\rangle_{\rm REN} \sim
\frac{\hbar}{4 \pi^2} \frac{H_0^4}{H}
+ 2 \epsilon H \langle\varphi^2\rangle_{\rm REN} = 
\frac{\hbar}{4 \pi^2} \frac{H_0^4}{H}
+ \frac{2}{3} \frac{m^2}{H} \langle\varphi^2\rangle_{\rm REN} \,.
\label{timeder}
\ee
We note that this result is very different from the de Sitter result 
($H$ constant in time and $\epsilon = 0$) since 
the second term on the right hand side may dominate for large times. 

The above result in Eq. (\ref{phi2_rinormalizzato_leading}) is 
reminiscent~\footnote{We note that this resemblance holds on 
only retaining the linear terms in $\log a$.} of the usual 
Hartree-Fock 
term formulae for stochastic inflation \cite{stochastic} (see Eq. (5) of 
\cite{SY}):
\be
\frac{d}{d t} \langle\varphi^2\rangle_{\rm REN} =
\frac{\hbar}{4 \pi^2} H^3 - 2 \frac{m^2}{3 H} 
\langle\varphi^2\rangle_{\rm REN} \,.
\label{stochastic}
\ee
However, Eqs. (\ref{timeder}) and (\ref{stochastic}) differ not only in 
the driving terms, but in particular in the mass term: in our result there 
is a negative mass term, whereas in the stochastic approach in rigid
de Sitter space time there is a positive mass term.
Our result (\ref{timeder}) is completely consistent 
with Eq. (\ref{EQMPF_k}) where the gravitational terms change the sign of the 
mass term in the evolution equation for the fluctuations.
The difference is therefore due to the inclusion of 
gravitational fluctuations which are not properly taken into account in 
stochastic inflation. 

At the end of inflation, for the characteristic time scale of
slow-rollover $\delta t \sim 3 H_0 / m^2$, the correlator
in the expression in (\ref{phi2_rinormalizzato_cap5})
saturates to
\be
\langle\varphi^2\rangle_{\rm REN}^{\rm MAX}    \sim
\frac{\hbar}{4 \pi^2} H^2 \left (\frac{H_0}{H}\right )^4 
\frac{1}{6 \epsilon_0} =
\frac{\hbar}{8 \pi^2} \frac{H_0^6}{H^2}\frac{1}{m^2}
\label{phi2_saturated}
\,.
\ee

It is instructive to also compute
$\langle \varphi \dot \varphi \rangle_{\rm REN}$ and
$\langle  \dot{\varphi}^2 \rangle_{\rm REN}$
by the same method, using
the approximate relation for the infrared modes
$\dot{\varphi}_k \simeq -(\dot{H}/H) \varphi_k$.
The final result (omitting all the steps in Eqs.
(\ref{phi2_UV},\ref{phi2_IR},\ref{phi2_adiabatico_5},
\ref{phi2_rinormalizzato_cap5})) is:
\be
H \langle \varphi \dot \varphi \rangle_{\rm REN} =
H \langle \varphi \dot \varphi \rangle^{IR}+\frac{\hbar}{16 \pi^2}
H^4 \left[D+ E \frac{m^2}{H^2} +\frac{8}{3} \frac{m^2}{H^2}
  \ln{\left(\frac{\nu H}{\lambda m}\right)} + 4 +\frac{4}{9} \frac{m^2}{H^2}+
{\cal O}\left( \frac{1}{a^3}\right) \right] \,
\label{phidotphi}
\ee
\be
 \langle \dot{\varphi}^2 \rangle_{\rm REN} = \langle \dot{\varphi}^2
\rangle^{IR}+\frac{\hbar}{16 \pi^2}
H^4 \left[F+ G \frac{m^2}{H^2} +2 \frac{m^2}{H^2}
  \ln{\left(\frac{\nu H}{\lambda m}\right)} + \frac{61}{60} -\frac{27}{10} \frac{m^2}{H^2}+
{\cal O}\left( \frac{1}{a^3}\right) \right] \, ,
\label{phidotsquare}
\ee
where $D$, $E$, $F$ and $G$ may be written in terms of generalized
hypergeometric functions and for the infrared contribution one
has~\footnote{More precisely, one may obtain the infrared
  contribution from the behaviour of the Mukhanov variable,
$ \langle \varphi \dot \varphi \rangle^{IR} 
\simeq \left[ H \epsilon +\ddot{H}/(2\dot{H})\right]
\langle \varphi^2 \rangle^{IR}$, where the second term is of order
${\cal O}(\epsilon^2)$.}
\be
\langle \varphi \dot \varphi \rangle^{IR} 
\simeq H \epsilon \langle \varphi^2 \rangle^{IR} \, .
\label{phidotphi_ir}
\ee
\be
\langle \dot{\varphi}^2 \rangle^{IR} 
\simeq H^2 \epsilon^2 \langle \varphi^2 \rangle^{IR} \, .
\label{phidotsquare_ir}
\ee


\section{Approximate Solution for Second Order Scalar Perturbations}
\label{seven}
An approximate solution for the leading quantities to second order
can be obtained for large scales where the infrared modes dominate.
Let us consider $\varphi^{(2)}$ which satisfies Eq.
(\ref{Eq_mukhanov_second}).
On neglecting terms ${\cal O} (\ddot H/H^3, \dot H^2/H^4)$ and the second
order derivative we obtain:
\be
3 H \dot \varphi^{(2)} - m^2 \varphi^{(2)} \simeq - V_\phi \frac{m^2}{6 H^2}
\frac{\varphi^2}{M_{\rm pl}^2} \,,
\ee
which can be rewritten, on also 
using the homogeneous equation of motion for the inflaton, as:
\be
3 \frac{d}{dt}\left( H \varphi^{(2)} \right) \simeq
\dot{\phi} \frac{m^2}{2 H} 
\frac{\varphi^2}{M_{\rm pl}^2} \,.
\label{approx_mukhanov_second}
\ee
Let us note that we use the notation $ M_{\rm pl}^2=1/(8\pi G)$ for
the (reduced) Planck mass definition. In the same approximation the energy 
constraint is:
\be
2 \alpha^{(2)} (3 H^2 + \dot H) \simeq - 6 \frac{\dot H}{M_{\rm pl}^2}
\varphi^2 + \frac{1}{M_{\rm pl}^2} \left[ - \dot \phi \dot \varphi^{(2)} -
V_\phi \varphi^{(2)} - \frac{1}{2} V_{\phi \phi} \varphi^2 \right] \,.
\ee

The expression we want is $\langle \varphi^{(2)} \rangle$,
which is averaged over the initial vacuum state, and can be obtained 
by using the leading contribution in
Eq. (\ref{phi2_rinormalizzato_leading})
to the quadratic correlator of the first order fluctuations
$\langle\varphi^2\rangle_{\rm REN}$.
We obtain directly from the last equation:
\be
\langle \varphi^{(2)}  \rangle \simeq
\frac{\hbar}{24 \pi^2} \dot \phi \frac{H_0^2 m^2}{M_{\rm 
pl}^2 \epsilon_0^2} \left[ \frac{1}{4 H^3} + \frac{1}{2 H H_0^2} 
\ln \frac{H}{M}
\right]
\ee
where $M$ is a constant with the dimensions of a mass.
It is also convenient to exhibit a slightly different form, which is of 
course equivalent in the large $a$ limit at the end of inflation since the
infrared contributions are then a maximum.

Since $\dot{\phi}$ is almost constant one may also obtain,
from Eq. (\ref{approx_mukhanov_second}),
\be
\langle \varphi^{(2)} \rangle^{\rm MAX} \simeq 
\frac{\dot{\phi}}{4H M_{\rm pl}^2} 
\langle\varphi^2\rangle_{\rm REN}^{\rm MAX} \,.
\label{phi2_ren_leading}
\ee
It is also useful to write the contribution to
$\langle \alpha^{(2)}  \rangle$ which we shall later use to study the
backreaction effects on some scalar observables.
One obtains for its leading behaviour
\be
\langle \alpha^{(2)} \rangle^{\rm MAX} \simeq \frac{\dot{\phi}}{2 H M_{\rm 
pl}^2}
\langle \varphi^{(2)} \rangle^{\rm MAX} -\frac{3}{4}\frac{1}{M_{\rm pl}^2}
\frac{\dot{H}}{H^2}
\langle\varphi^2\rangle_{\rm REN}^{\rm MAX}
\simeq  \frac{\epsilon}{M_{\rm pl}^2}
\langle\varphi^2\rangle_{\rm REN}^{\rm MAX}
\label{alpha2_ren_leading}
\ee


\section{Approaches to the back-reaction}\label{eight}
One may follow different approaches in order to study the backreaction 
effects due to cosmological fluctuations.
In the following we shall consider two methods in order 
to tackle this issue.

One consists of considering only the first order perturbations,
then imposing to first order the energy and momentum constraints
and finally defining an effective EMT by including all the quadratic terms
present in the Einstein equations. Subsequently one averages everything
over the quantum vacuum and first order quantities disappear.
One finds that the effective EMT which appears in the averaged
Einstein equations is modified by the back-reaction. The results obtained 
in sections \ref{two}, \ref{five} and \ref{six} are sufficient for
this purpose. 
However one still has to study how any observable averaged over the
quantum vacuum is then affected.

A second approach is related to the standard perturbation analysis of
the Einstein equations up to second order. In this case we impose the
energy and momentum constraints and study the inflaton equation of
motion perturbatively up to second order.
One does not define any modified EMT but directly studies in this
framework any observable averaged over the quantum vacuum.
In order to pursue this approach results given in sections
\ref{three}, \ref{four} and \ref{seven} are also necessary. 
\subsection{The Energy-Momentum Tensor of Cosmological Fluctuations}
\label{eight1}
By the EMT of cosmological fluctuations we mean the second order part
both of the scalar field EMT {\em and} of the Einstein 
tensor which is quadratic in first order fluctuations:
\be
\tau^\mu_\nu \equiv T^{\mu \, {\rm quadratic}}_\nu - M_{\rm pl}^2 
G^{\mu \, {\rm quadratic}}_\nu \, .
\ee
This method of considering the EMT of gravitational fluctuations is 
treated in textbooks \cite{weinberg} and has also been 
previously used in \cite{ABML,ABMP,AF}.
In this scheme one considers the modified Einstein equations
$M_{\rm pl}^2 G^{\mu\, (0)}_\nu = T^{\mu\, (0)}_\nu +
\langle \tau^\mu_\nu \rangle$
which therefore include back-reaction effects.

Let us consider the EMT of cosmological fluctuations averaged over the 
state annihiled by the operator $\hat b$ defined in
Eq. (\ref{quantumFourier}). 

For a generic potential the leading terms in the energy density are
\be
\langle \tau^0_0 \rangle \sim -
\frac{V_{\phi \phi}}{2} \langle \varphi^2 \rangle + 12 H^2 M_{\rm pl}^2 
\langle \alpha^2 \rangle = -
\frac{V_{\phi \phi}}{2} \langle \varphi^2 \rangle - 6 \dot H \langle 
\varphi^2 \rangle
\label{leading}
\ee

It is important to note that the second term on the right hand side is 
in general {\em positive} and {\em larger} than the first during 
inflation. This
second term is the contribution of metric perturbations. On using the 
slow-rollover parameters $\epsilon$ and $\eta$ ($ \equiv M_{\rm pl}^2 
\frac{V_{\phi \phi}}{V}$)
we can rewrite the leading terms in Eq. (\ref{leading}) during 
slow-rollover as
\be
\langle \tau^0_0 \rangle \equiv - \varepsilon 
\sim - \frac{V_{\phi \phi}}{2} \langle \varphi^2 \rangle \left( 1 
- 4 \frac{\epsilon}{\eta}
\right) \,.
\label{tauzerozero}
\ee

Analogously the average pressure is:
\be
\langle \tau^i_j \rangle \equiv p \, \delta^i_j \sim \delta^i_j \left( 
- \frac{V_{\phi \phi}}{2} \langle \varphi^2 \rangle + 12 H^2 M_{\rm pl}^2
\langle \alpha^2 \rangle \right) \sim - \varepsilon \, \delta^i_j 
\ee

We now restrict the analysis to a quadratic potential
with $\eta \simeq \epsilon$. On using 
Eq. (\ref{phi2_rinormalizzato_leading}) the leading behaviour for the
initial time of the energy and pressure density is:
\be
\varepsilon_{\rm REN} \sim - p_{\rm REN}
\sim - \frac{3}{2} m^2 \langle\varphi^2\rangle_{\rm REN} \sim -
\frac{3 \hbar}{8 \pi^2} m^2 H^2 \left (\frac{H_0}{H}\right )^4
\ln a = - \frac{9 \hbar}{8 \pi^2} \epsilon H_0^4 \ln a
\,,
\label{pippo}
\ee
further, on using Eq.(\ref{phi2_saturated}), 
we have the following maximum value:
\be
\varepsilon_{\rm REN}^{\rm MAX} \sim - p_{\rm REN}^{\rm MAX}
\sim -  \frac{3 \hbar}{16 \pi^2}
\frac{H_0^6}{H^2}
\ee

we then obtain that for 
\be
H_0 \sim (16 \pi^2/\hbar)^{1/6} H^{2/3} M_{\rm pl}^{1/3} \,.
\ee
the EMT tensor of cosmological fluctuations cannot be neglected with 
respect to the energy content of the homogeneous classical inflaton.

As a final remark we note that the amplitude and time dependence of the 
EMT depends on the approximation used for linear perturbations. The 
slightly different results in the amplitude of the EMT with respect to 
\cite{ABML,ABMP} are explained in Appendix C.
\subsection{Back-Reaction on the Geometry}\label{eight2}
In the perturbative approach to the Einstein equations any
back-reaction effect is analyzed by evaluating perturbatively
quantities which characterize the geometry.

The expansion scalar is defined as:
\be
\Theta = \nabla_\mu u^\mu \,,
\ee
where $u^\mu$ is a four-vector field defining the comoving frame. The 
four-vector is normalized, $u_\mu u^\mu = - 1$. To second order the four 
vector is:
\be
u^0 = 1 - \alpha + \frac{3}{2} \alpha^2 
- \alpha^{(2)} \,, \quad u^i = \frac{\beta^{,i} +
\beta^{(2)\,,i}}{a} 
\ee

and the expansion scalar reads:
\begin{eqnarray}
\Theta &=& \Theta^{(0)}+\Theta^{(1)}+\Theta^{(2)}
\nonumber \\
&=& 3 H -3 H \alpha +\frac{1}{a} \nabla^2 \beta -3 H \alpha^{(2)} 
+ \frac{1}{a} \nabla^2 \beta^{(2)} -
\nonumber \\
& &
+\frac{9}{2} H \alpha^2 + \frac{1}{a}
 \vec{\nabla}\alpha \cdot \vec{\nabla} \beta + 
\frac{1}{4}\vec{\nabla}\beta
\cdot \vec{\nabla}
 \dot{\beta} \, .
\end{eqnarray}

The leading terms, once we average, are:
\begin{eqnarray}
\langle \Theta \rangle &=& \Theta^{(0)} + \langle \Theta^{(2)} \rangle
\nonumber \\
&=& 3 H \left( 1 + \frac{3}{2} \langle \alpha^2 \rangle - \langle 
\alpha^{(2)} \rangle \right)
\nonumber \\
&\simeq & 3 H \left( 1 - \epsilon
\frac{\langle \varphi^2 \rangle^{\rm MAX}}{4 M_{\rm pl}^2}
\right)
\label{thetaresult}
\end{eqnarray}
where the results given in Eqs. (\ref{phi2_rinormalizzato_leading}) and 
(\ref{alpha2_ren_leading}) have been used, together with the relations 
$\alpha=\dot{\phi}/(2H M_{\rm pl}^2) \varphi$ and
$\dot{\phi}^2=-2\dot{H} M_{\rm pl}^2$, in order to calculate
the only two non negligible second order contributions.
We note that by leading terms we mean the first corrections in
$\epsilon$ which are also leading in $\log a$, thus essentially including 
the infrared contribution present in the renormalized quantities,
as already discussed.  
We observe that the result in Eq. (\ref{thetaresult}) is completely 
consistent with the considerations on the EMT made in the previous 
subsection. When the magnitude of the EMT of cosmological fluctuations is 
of the same order as that of the background $\langle \Theta \rangle$ 
changes significantly.

The result in Eq. (\ref{thetaresult}) is in contrast with \cite{BRAND}, 
where a vanishing result was obtained for the back-reaction on the 
expansion rate $\langle \Theta \rangle$ and where only first order
classical perturbations are considered. 
We note that also in our gauge choice $\Theta$ is related to the Einstein 
tensor for long wavelenghts and leading order in $\epsilon$:
\be
\Theta \simeq
3H \left(1-\alpha -\alpha^{(2)} +\frac{3}{2} \alpha^2 \right) 
\simeq \sqrt{-3 G_0^0}\,,
\ee
and therefore the result in Eq. (\ref{thetaresult}) could also be obtained by 
expanding $\sqrt{-24 \pi G T_0^0}$ to second order and averaging over the
quantum vacuum. The difference with respect to 
\cite{BRAND} may be due to the absence of second order fluctuations in 
\cite{BRAND} and/or to a peculiarity of inflation driven by a quadratic 
potential.

We now consider another geometric scalar observable which is associated
with the rate of change of the expansion scalar, i.e.:
\be
\Omega = u^\nu \nabla_\nu \nabla_\mu u^\mu \,.
\ee 
Its expansion up to second order is given by
\begin{eqnarray}
\Omega &=& \Omega^{(0)}+\Omega^{(1)}+\Omega^{(2)}
\nonumber \\
&=& 3 \dot{H} - 6 \dot{H} \alpha -3 H \dot{\alpha} -\frac{H}{a} \nabla^2
\beta
 +\frac{1}{a} \nabla^2 \dot{\beta} +
\nonumber \\
& &
- 6 \dot{H} \alpha^{(2)} -3 H \dot{\alpha}^{(2)} -\frac{H}{a} \nabla^2
\beta^{(2)} +\frac{1}{a} \nabla^2 \dot{\beta}^{(2)}
\nonumber \\
& &
+12 \dot{H} \alpha^2 +12 H \alpha \dot{\alpha}-4 \frac{H}{a}
 \vec{\nabla}\alpha \cdot \vec{\nabla} \beta +\frac{H}{a}
 \alpha \nabla^2 \beta -
\nonumber \\
& &
 - \frac{1}{a}\alpha \nabla^2 \dot{\beta}+ \frac{1}{4} \frac{d}{dt}
\left (\vec{\nabla}\beta \cdot \vec{\nabla} \dot{\beta} \right ) +
\frac{1}{a} \frac{d}{dt}\left (\vec{\nabla
}\alpha \cdot \vec{\nabla} \beta
\right )+
\nonumber \\ & &
+ \frac{1}{a^2} \vec{\nabla}\beta \cdot \vec{\nabla}\left ( \nabla^2\beta
\right )
\end{eqnarray}

For the homogeneous case $\Theta = 3 H$ and therefore
$\Omega = 3 \dot H$. During slow-rollover for a massive inflaton $\dot H
$ ($\simeq - m^2/3$) is constant in time, and therefore is a 
gauge-invariant quantity up to second order \cite{SW,bruni}.

The leading terms, once we average, are:
\begin{eqnarray}
\langle \Omega \rangle &=& \Omega^{(0)} + \langle \Omega^{(2)} \rangle
\nonumber \\
&=& 3 \dot{H} 
- 6 \dot{H} \langle \alpha^{(2)} \rangle
-3 H \langle \dot{\alpha}^{(2)} \rangle
+12 \dot{H} \langle \alpha^2 \rangle + 12 H \langle \alpha \dot{\alpha} 
\rangle \nonumber \\
& \simeq & 3 \dot H \left( 1 + F(\epsilon,\log a) \right)
\end{eqnarray}
where $F(\epsilon,\log a) \simeq \epsilon F_1(H) + \epsilon^2 \log a \,
F_2(H)+ \cdots $.
Therefore one observes that in this case the leading corrections
to order $\epsilon$ which come from the infrared region cancel,
while these were present in $\langle \Theta \rangle$.
Let us again note that at the end of inflation driven by a  massive
inflaton $\log a \simeq 1/(2\epsilon_0)$  so that any $\log a$
correction effectively reduces the perturbative order in powers
of $\epsilon$ of the correction.
 
The result on $\langle \Omega \rangle$ 
should be interpreted with care since we are not able to go to
higher order self-consistently within our approximation.
We also note that this vanishing result may also be a peculiarity of
inflation driven by a quadratic potential. In any case the rate of change
of the Hubble parameter can also be studied more carefully at the
background level, without considering any back-reaction.
Including the first non trivial corrections one finds
\be
\dot H \simeq - \frac{m^2}{3} \left( 1 - \frac{1}{9}\frac{m^2}{H^2} \right) \,.
\ee

To conclude this section we stress that in both the results (\ref{pippo}) and 
(\ref{thetaresult}) the corrections to the background 
values, which grow with time, are ${\cal O} (\epsilon)$. In the de Sitter limit ($\epsilon=0$) 
these corrections vanish, consistently with the absence of physical scalar 
perturbations to first order for a universe driven by a cosmological 
constant.


\section{Discussion and Conclusions}\label{nine}
The renormalized EMT of cosmological fluctuations during 
inflation with a quadratic potential is studied. With respect to our 
previous work 
\cite{paperI} we have  
self-consistently taken into account the gravitational fluctuations 
accompanying the scalar field fluctuations. 

We find that the renormalized EMT of cosmological fluctuations 
during slow-rollover carries 
negative energy density and has a de Sitter like equation of state to 
leading order. The negative sign for the renormalized energy density is 
due to the inclusion of gravitational fluctuations 
(see also \cite{ABML,ABMP} for the same claim regarding long-wavelength 
modes) and did not appear in the previous calculation in rigid space-time 
\cite{paperI}. This effect is generally true for single field inflationary models with 
$\epsilon > \eta/4$, as is seen from Eq. (\ref{tauzerozero}).

The back-reaction problem was also treated up to second-order 
in perturbation theory. For this purpose we extended the UCG gauge beyond the
linear order by taking into account vector and tensor perturbations to second 
order. The inclusion of vector and tensor perturbations to second order is 
important in order to obtain the correct equations for the second order 
scalar perturbations. We derived the equation
for the second order scalar field fluctuations and approximatively solved the 
system of second order equations for single field inflation during 
slow-rollover. 
We computed the expansion rate $\Theta$ and its 
coordinate invariant derivative $\Omega$ to second order, and we found 
that the expansion rate is decreased by the EMT of cosmological
fluctuations, while $\Omega$ is not affected to leading order.
A non vanishing correction for $\Omega$ will probably appear 
in the next order.

We do not find evidence that the back-reaction of cosmological 
fluctuations is a gauge artifact as is claimed in \cite{UNRUH}. The 
second order corrections to the local expansion rate $\Theta$ are
non vanishing and are computed to lowest order in Eq. (\ref{thetaresult}). 
We note that the back-reaction of cosmological scalar fluctuations 
vanishes in the de Sitter limit ($\epsilon=0$), as it should.
Since $\Theta$ is not a gauge-invariant quantity in the model 
under consideration~\cite{bruni,SW}, we also computed 
$\Theta$ in the gauge wherein the inflaton is a clock: such a check
was done by a completely new calculation (illustrated in appendix
C) in order to also study the transformation properties of the
metric elements.  
We confirm  that back-reaction is an ${\cal O} (\epsilon)$ effect,
but its time behavior is  different (there is no $\log a$ enhancement from
the infrared region) because of the differing times in the two frames.
In this new gauge calculations are also more involved since a
higher accuracy in the solution for the Mukhanov variable would be required. 

It is interesting to note that the inclusion of gravitational fluctuations 
sistematically contributes negatively. To first order 
gravitational fluctuations contribute 
negatively to the effective mass for the fluctuations, and to second order 
carry negative energy density. The self-consistent inclusion of gravity 
also changes the stochastic picture in an inflationary background.

The approach we have used is mean field theory on a curved space-time. 
The procedure of averaging over a quantum state with a cut-off 
related to the patch which undergoes inflation can be seen as a 
spatial average over the particle horizon. This average is completely 
different from the average over the Hubble region during inflation used in 
the stochastic approach \cite{stochastic} and 
in which gravitational fluctuations 
were neglected. We think that the 
difference between the results of stochastic inflation and those found 
here are not just due to a different coarse graining, but are also due 
to our inclusion of 
gravitational fluctuations. 
We think that the relation between these two different 
approaches should be further investigated. 

    \vspace{0.5cm}
    \centerline{\bf Acknowledgments}
    \vspace{0.2cm}

We wish to thank A. A. Starobinsky for discussions and suggestions 
on the draft. We also wish to thank R. Abramo, R. Brandenberger, A. 
Linde and S. Matarrese for comments. F. F. thanks R. Abramo and 
R. Brandenberger for many conversations on the back-reaction problem over 
the previous years and N. Bartolo and B. Losic for discussions.


\section{Appendix A: Comparison with WKB methods}
In this appendix we show how the procedure used in this paper
to match solutions which are good approximations in different ranges of
$k$ space includes the recent application of the WKB method in the context 
of cosmological perturbations \cite{habib,ms}. In this paper we have
chosen a Bessel form in Eq. (\ref{ultraviolet}) for modes which are still
inside the Hubble radius.
The scaled equation ($\psi_k=a^{3/2}\varphi_k$) associated with 
eq. (\ref{EQMPF_k}) is given by

\be
\ddot{\psi}_k +\left( \left(\frac{k}{a H \nu'} \right)^2 -1\right)H^2
  \nu'^2 \psi_k=0\, ,
\label{eq1}
\ee
where $\nu'=\nu/\lambda$.
The turning-point of this equation is obtained for
\be
a(t)=\frac{k}{H(t) \nu'(t)}\,.
\label{T_P_App_A}
\ee
So, for our calculation, we can neglect the weak time dependence
on the right side of Eq.(\ref{T_P_App_A}) with respect the exponential dependence
on the left side and obtain the following result for the turning-point
\be
t_k \simeq -\frac{H_0}{\dot{H}_0}
\left(1- \sqrt{1+2\frac{\dot{H}_0}{H_0^2} \log{\frac{\lambda_0 k}{H_0 
\nu_0}}} \,
\right)\simeq \frac{1}{H_0} \log{\frac{\lambda_0 k}{H_0 \nu_0}}\,.
\ee
The solution near the turning-point of Eq. (\ref{EQMPF_k}), on using the WKB 
method, is given by
\be
\varphi_k=\frac{1}{a^{3/2}}\left [B_1 \, Ai\left(\alpha_k^{1/3} (t-
  t_k) \right)+
B_2 \, Bi\left(\alpha_k^{1/3} (t- t_k)\right) \right ]. \label{WKBab}
\ee
Around the turning-point we can consider the following approximate 
expression~\cite{gradsht} in Eq. (\ref{ultraviolet})
\be
H_\nu^{(1)}\left(x\right) =  \frac{w}{\sqrt{3}}
e^{i \left[ \frac{\pi}{6}+\nu \left(w
    -\frac{w^3}{3}-\arctan{w}\right)\right]}
H^{(1)}_{1/3}\left(\frac{\nu}{3}w^3\right) +
{\cal O}\left(\frac{1}{\nu}\right) 
\ee
or 
\be
H_\nu^{(1)}\left(x\right) \approx \left(\frac{2}{\nu} 
\right)^{\frac{1}{3}}
e^{i \nu \left(w -\frac{w^3}{3}-\arctan{w}\right)}
\left\{ Ai\left[ -\left(\frac{\nu}{2}\right)^{\frac{2}{3}} w^2 \right]
-i \, Bi\left[ -\left(\frac{\nu}{2}\right)^{\frac{2}{3}} w^2
  \right]\right\} \, ,
\ee
where $w=\frac{1}{\nu}\sqrt{x^2-\nu^2}$. 
We have checked numerically that the above approximations are remarkably good. 
The Airy functions are precisely the 
functions used to match the WKB solution, eq.(\ref{WKBab}), inside and
outside the Hubble radius \cite{ms}.
In particular around the turning-point the match is specified by
\be
-\left(\frac{\nu(t_k)}{2}\right)^{\frac{2}{3}}w^2 \approx
\alpha_k^{1/3} \left(t-t_k\right) \, .
\ee
Therefore, by choosing a Bessel form inside the
Hubble radius one already has the correct form in the vicinity of the 
turning point. This cannot be achieved with the WKB approximation, since 
the WKB solution is singular in correspondence of the turning point.


\section{Appendix B: The fourth order adiabatic expansion}
In order to remove the divergences which appear in the integrated
quantities as poles in the $\Gamma$ functions, we shall employ the method
of {\em adiabatic subtraction} \cite{adiabatic}. Such a method consists of
replacing our functions with an expansion in powers of derivatives of the
logarithm of the scale factor.
This expansion
coincides with the adiabatic expansion introduced by Lewis in \cite{lewis}
for a time dependent oscillator.

Usually it is more convenient to formulate the adiabatic expansion by
using the modulus of mode functions $x_k =
|\varphi_{\bf k}/\sqrt{2}|$ and
the conformal time $\eta$ \cite{adiabatic} ($d \eta = d t/a $).
We follow this procedure and write an expansion in derivatives with
respect to the conformal time (denoted by $'$) for $x_k$. We then go back 
to cosmic time and insert the expansion in the expectation values 
we wish to compute.  Adiabatic expansions in cosmic time and
conformal time lead to equivalent results, because of
the explicit covariance under time reparametrization \cite{HMPM}.

The variable $x_k$ satisfies
the following Pinney equation:

\be
\ddot{x}_k+3 H \dot{x}_k+
\left [\frac{k^2}{a^2}+m^2+6 \dot{H} \right ] x_k=\frac{1}{a^6 x^3_k} \,.
\label{EQMPF_k_x}
\ee

Following \cite{paperI} we rewrite Eq. (\ref{EQMPF_k_x}) in conformal
time in the following
way:
\be
(a x_k)'' + \Omega_k^2 \, (a  x_k) =
\frac{1}{(a x_k)^3}
\label{conformal_eq_x_k}
\ee
where
\be
\Omega_k^2 = k^2 + m^2 a^2 - \frac{1}{6} a^2 \tilde{R}
\label{conformal_freq_x_k}
\ee
and $\tilde{R}$ is:
\be
\tilde{R} = R - 36 \dot{H} \,.
\label{ricci_art2}
\ee
with $R=6\frac{a''}{a^3}$ the Ricci curvature.
From Eqs.
(\ref{conformal_eq_x_k},\ref{conformal_freq_x_k},\ref{ricci_art2})
one obtains the expansion for $x_k$ up to the fourth adiabatic order:
\be
x_{k}^{(4)}=\frac{1}{a}\frac{1}{\Omega_k^{1/2}}
\left( 1-\frac{1}{4} \epsilon_2+\frac{5}{32}\epsilon_2^2-
\frac{1}{4}\epsilon_4 \right)
\label{fourth_1}
\ee
where $\Omega_k$ is defined in Eq. (\ref{conformal_freq_x_k}) and
$\epsilon_2 \,, \epsilon_4$ are given by:
\begin{eqnarray}
\epsilon_2&=&-\frac{1}{2}\frac{\Omega_k^{''}}{\Omega_k^3}+\frac{3}{4}
\frac{\Omega_k^{'2}}{\Omega_k^4} \nonumber \\
\epsilon_4 &=& \frac{1}{4}\frac{\Omega_k^{'}}{\Omega_k^3}\epsilon_2^{'}-
\frac{1}{4}\frac{1}{\Omega_k^2}\epsilon_2^{''}
\end{eqnarray}

The solution in Eq. (\ref{fourth_1}) must be expanded again since $\tilde{R}$
is of adiabatic order 2. Therefore $x_k^{(4)}$ is:

\begin{eqnarray}
x_{k (4)} &=& \frac{1}{c^{1/2}}\frac{1}{\Sigma_k^{1/2}} \Bigl\{
1+\frac{1}{4}c\frac{\tilde{R}}{6}\frac{1}{\Sigma_k^2}+\frac{5}{32}c^2
\frac{\tilde{R}^2}{36} \frac{1}{\Sigma_k^4} \nonumber \\
 & & +\frac{1}{16}\frac{1}{\Sigma_k^4}\left[c^{''} \left(m^2-
\frac{\tilde{R}}{6}
 \right)-2 c^{'}
 \frac{\tilde{R}^{'}}{6}-c\frac{\tilde{R}^{''}}{6}\right]
\nonumber \\
 & &  - \frac{5}{64}\frac{1}{\Sigma_k^6}\left[ c^{'2}m^4-2c^{'2}m^2
 \frac{\tilde{R}}{6}-2 c^{'}m^2c\frac{\tilde{R}^{'}}{6}\right] \nonumber \\
 & & + \frac{9}{64}\frac{1}{\Sigma_k^6} c \frac{\tilde{R}}{6}
c^{''}m^2-
\frac{65}{256}  \frac{1}{\Sigma_k^8} c \frac{\tilde{R}}{6} c^{'2}m^4+
\frac{5}{32}\epsilon_{2*}^2- \frac{1}{4}\epsilon_{4*}   \Bigr\}
\label{EA4}
\end{eqnarray}

where $c=a^2$ and
\begin{eqnarray}
\Sigma_k&=&(k^2 + a^2 m^2)^{1/2} \nonumber \\
\epsilon_{2*}&=&-\frac{1}{2}\frac{\Sigma_k^{''}}{\Sigma_k^3}+\frac{3}{4}
\frac{\Sigma_k^{'2}}{\Sigma_k^4} \nonumber \\
\epsilon_{4*} &=&
\frac{1}{4}\frac{\Sigma_k^{'}}{\Sigma_k^3}\epsilon_2^{'}-
\frac{1}{4}\frac{1}{\Sigma_k^2}\epsilon_2^{''}
\end{eqnarray}

\section{Appendix C: Comparison with calculations in different gauges}

{\bf a.} In \cite{ABML,ABMP} a different procedure was followed. 
The calculation
was performed in the Newtonian gauge (see Eq. (\ref{longitudinal}))and
the regularization of the metric perturbation $\langle \Phi^2 \rangle$
was considered. We can compare our
results by using the relation between field fluctuations $\varphi$
in the uniform curvature gauge and field fluctuations $\varphi_N$ in the
Newtonian gauge:
\be
\varphi = \varphi_N + \frac{\dot \phi}{H} \Phi \,.
\ee
Assuming that $\dot \Phi$ in negligible on large scales, we obtain
\cite{MFB}:
\be
\varphi \simeq \frac{\dot \phi}{H}
\left( \frac{\epsilon - 1}{\epsilon} \right) \Phi \,,
\ee
which, on using the approximation used in \cite{MFB,ABML,ABMP}, leads to
\be
\varphi_k \simeq \frac{\dot \phi}{H}
\left( \frac{\epsilon - 1}{\epsilon} \right)
k^{-3/2} \frac{m}{2 \sqrt{6}  M_{\rm pl}}
\left[ 1 - \frac{\log [ k/(a H) ]}{
1+\log [ a_{\rm end}/a ]} \right] \,,
\ee
where $a_{\rm end}$ is the scale factor at the end of inflation. This 
approximation for the gauge-invariant field fluctuations is different from 
our Eq. (\ref{solution_infrarossa_in}).

\vspace{1cm}

{\bf b.} After this paper appeared in the electronic arXiv, 
it was suggested that corrections to $< \Theta >$ should disappear in
the gauge wherein 
the inflaton is a clock, homogeneous in space. 
Let us note that the inflaton can work as a clock only when it is not
oscillating, as is the case in the slow-rollover regime. 

In order to check this claim we performed a new calculation,
since the gauge where the inflaton is homogeneous is different to 
the one used in this paper. The gauge in which the inflaton is 
unperturbed is described by a metric with three scalar degrees of
freedom and, again, this is a choice which fixes the gauge completely
because of the  presence of the scalar field.
As before we do not pay attention to the vector and tensor degrees of
freedom and therefore we consider the following second order line element
depending only on three scalars:
\be
ds^2=-(1+2 \tilde{\alpha}+2\tilde{\alpha}^{(2)} )d\tilde{t}^2-
a (\tilde{\beta}_{,i}+
\tilde{\beta}^{(2)}_{,i}) d\tilde{t} d\tilde{x}^i+
a^2 \left [ 
\delta_{ij}(1 - 2 \tilde{\psi}-2 \tilde{\psi}^{(2)})\right ]
d\tilde{x}^i d\tilde{x}^j \,,
\label{PER_HOMO}
\ee
where we have taken $\tilde a ( \tilde t ) = a ( \tilde t )$,
since in the perturbative approach $a ( \tilde t )$ and
$\tilde a (\tilde t )$ satisfy the same equation in $t$ and $\tilde
t$, respectively  
(however the scale factor is not observable for the case under 
consideration). 
We expect that $\tilde{\psi}$ will satisfy a dynamical equation of
motion in this gauge.

Instead of solving the system in this gauge, we find solutions for the 
metric perturbations in (\ref{PER_HOMO})
by using its gauge relation with the metric in
(\ref{metric_second}).
An infinitesimal coordinate trasformation up to second order \cite{bruni}:
\be
x^\mu \rightarrow \tilde{x}^\mu= x^\mu + \epsilon^\mu_{(1)} +\frac{1}{2}
\left(\epsilon^{\mu}_{(1),\nu}
\epsilon^{\nu}_{(1)} + \epsilon^{\mu}_{(2)}\right) \,,
\label{Trasf_gauge_coord}
\ee
(where $\epsilon_{(1)}$ and $\epsilon_{(2)}$ are the coordinate changes to 
first an second order, respectively)
induces the most generic change in a geometric object $T = T^{(0)} + T^{(1)} 
+ T^{(2)}$:
\be
T^{(1)} \rightarrow \tilde{T}^{(1)}=
T^{(1)} - \mathcal{L}_{\epsilon_{(1)}} T^{(0)}
\label{Trasf_tens_ord1}
\ee
\be
T^{(2)} \rightarrow  \tilde{T}^{(2)}=
T^{(2)} - \mathcal{L}_{\epsilon_{(1)}} T^{(1)} + 
\frac{1}{2}
\left(\mathcal{L}_{\epsilon_{(1)}}^2 T_0 - 
\mathcal{L}_{\epsilon_{(2)}} T_0 \right)
\label{Trasf_tens_ord2}
\ee
The time reparametrization which relates the two gauges can be found by 
imposing that the field perturbation is zero to first and second order in 
the metric (\ref{PER_HOMO}). According to Eqs. 
(\ref{Trasf_tens_ord1},\ref{Trasf_tens_ord2}) the field transforms as:
\be
\varphi \rightarrow \tilde{\varphi}= \varphi - \epsilon^0_{(1)} \dot{\phi}
\label{Trasf_inf_ord_uno}
\ee
\be
\varphi^{(2)} \rightarrow \tilde{\varphi}^{(2)}= 
\varphi^{(2)} - \epsilon^0_{(1)} \dot{\varphi} + \frac{1}{2}
\left[\epsilon^0_{(1)} 
\left(\epsilon^0_{(1)}\dot{\phi}\right)^.
- \epsilon^0_{(2)} \dot{\phi}\right] \,,
\label{Trasf_inf_ord_due}
\ee
which leads to
\begin{eqnarray}
\epsilon^0_{(1)} &=& \frac{\varphi}{\dot{\phi}} = \frac{\zeta}{H} 
\\
\epsilon^0_{(2)} &=& \frac{2}{\dot{\phi}}\varphi^{(2)}
- \frac{1}{\dot{\phi}^2}\varphi \dot{\varphi} \simeq
-\frac{\zeta \dot{\zeta}}{H^2} -\frac{1}{2} \frac{\ddot{H}}{\dot{H}H^2}
  \zeta^2\,,
\label{coord_trans}
\end{eqnarray}
where we have used the gauge invariant curvature perturbation
$\zeta=H\varphi/\dot{\phi}$, which is constant on large scales and the
expression in eq. (\ref{phi2_ren_leading}).
From the above we understand that we need a large time
reparametrization in order to keep the scalar field homogeneous in space.

The above trasformation, up to second order, is not the one required
since it would lead to the presence of a scalar second order
contribution in the traceless part of the metric which depends on a
quadratic form in $\partial_i\beta$ and $\partial_i\epsilon^0_{(1)}$.
In order to kill this contribution a scalar second order transformation in the
spatial part of the coordinates is also required,
$\epsilon_{(2)}^i = \partial^i \epsilon^s_{(2)}$, while still having
$\epsilon_{(1)}^i =0$.
We find
\be
\epsilon^s_{(2)}= \frac{1}{2a} \beta \epsilon^0_{(1)} - 
\frac{1}{2a^2} (\epsilon^0_{(1)})^2+ 
\frac{3}{2}\frac{\partial^i \partial^j}{(\nabla^2)^2}
\left[ \frac{1}{a^2}\epsilon^0_{(1)} D_{ij} \epsilon^0_{(1)} -
 \frac{1}{2a} \left(
\epsilon^0_{(1)} D_{ij}\beta + \beta  D_{ij} \epsilon^0_{(1)}
\right) \right] \, ,
\label{spacetransf}
\ee
where $D_{ij}=\partial_i \partial_j -1/3 \, \nabla^2 \delta_{ij}$.

On using the gauge transformation (\ref{Trasf_tens_ord1}) 
we have the following relation for the metric fluctuations to first order:
\be
\alphat = \alpha - \dot \epsilon^0_{(1)} \,, \quad 
\betat = \beta - \frac{2}{a} \epsilon^0_{(1)} \,, 
\quad \tilde{\psi} = H \epsilon^0_{(1)} \,.
\ee
On solving for the metric fluctuations in the UCG gauge, using Eqs. 
(\ref{Eq_beta},\ref{Eq_alpha}) we have:
\be
\alpha = \epsilon \, \zeta  \,, \quad \quad 
\frac{\nabla^2 \beta}{a} = 2 \epsilon \dot \zeta \,,
\ee 
whereas in the unperturbed scalar field gauge we have:
\be
\tilde{\alpha} = - \frac{\dot \zeta}{H}  \,, \quad  
\frac{\nabla^2 \tilde{\beta}}{a} = 2 \epsilon \dot \zeta - 2 
\frac{\nabla^2 \zeta}{a H} \,, \quad 
\tilde{\psi} = \zeta \,.
\ee
From the last relations we see that the cost of keeping the scalar 
field homogeneous in space is to squeeze a large fluctuation in the metric, 
i.e. $\tilde{\psi}$. In the UCG gauge metric fluctuations are
suppressed with respect to $\zeta$, i.e.
${\cal O} (\epsilon \dot \zeta \,, \epsilon \, \zeta)$, 
whereas in the gauge (\ref{PER_HOMO}) they are not. To second order we have:
\begin{eqnarray}
\alphat^{(2)} & = & \alpha^{(2)} - \epsilon^0_{(1)} \dot{\alpha} -2 \alpha
\dot \epsilon^0_{(1)} - \frac{\dot \epsilon^0_{(2)}}{2}+\frac{1}{2}\epsilon^0_{(1)}
\ddot \epsilon^0_{(1)} + \dot \epsilon^{0 \,\, 2}_{(1)}
\nonumber \\
& = &  \alpha^{(2)} - 2 \epsilon^2 \zeta^2 -
\frac{\dot \epsilon^0_{(2)}}{2} + 
\frac{1}{2} 
\frac{\ddot{H}}{H^3} \zeta^2 + \frac{\dot{\zeta}^2}{H^2} + \frac{1}{2} 
\frac{\zeta \ddot{\zeta} } {H^2} \, ,
\end{eqnarray}
where, from eq. (\ref{alpha2_ren_leading}), one has
$\alpha^{(2)} \simeq 2\epsilon^2 \zeta^2$. Moreover we find for
$\beta^{(2)}$ a transformation such that:
\be
\tilde{\beta}^{(2)}= \beta^{(2)} -\frac{1}{a} \epsilon_{(2)}^0
+ a \dot{\epsilon}^s_{(2)} +\frac{1}{2a}\frac{d}{d t} (\epsilon_{(1)}^0)^2
+\frac{\partial^i}{\nabla^2}
\left[ \frac{2}{a} \partial_i\epsilon_{(1)}^0 \dot{\epsilon}_{(1)}^0
- \epsilon_{(1)}^0  \partial_i \dot{\beta}-
H \epsilon_{(1)}^0 \partial_i \beta - \dot{\epsilon}_{(1)}^0 \partial_i
\beta - \frac{4}{a} \alpha \partial_i \epsilon_{(1)}^0 
\right]
\, .
\ee
Let us remember that we chose not to exhibit in detail the vector and tensor
degrees of freedom, but it is straightforward to do so.
It is suffices to say that to kill the possible vector
degrees of freedom arising in $\tilde{g}_{0i}$, a non vanishing vector 
component in the coordinate tranformation to second order, 
$\epsilon_{(2)}^{\perp \, i}$,
has to be introduced, but this does not affect the scalars to second order,
$\alphat^{(2)}$, $\betat^{(2)}$ and $\tilde{\psi}^{(2)}$.

Finally, for this last scalar we find
\begin{eqnarray}
\tilde{\psi}^{(2)} &=& \frac{H}{2}\epsilon^0_{(2)} - 
\frac{H}{2} \epsilon^0_{(1)} \dot \epsilon^0_{(1)} - 
\frac{\epsilon^{0 \, 2}_{(1)}}{2} \left( \dot H + 2 H^2 \right) +
\frac{1}{6a^2} \partial^i \epsilon^0_{(1)} \, \partial_i \epsilon^0_{(1)}
-\frac{1}{6a} \partial^i \beta \partial_i \epsilon^0_{(1)}
+\frac{1}{6} \nabla^2 \epsilon^s_{(2)}\nonumber \\
&=& \frac{H}{2}\epsilon^0_{(2)} -\zeta^2 -\frac{1}{2 H} \zeta
\dot{\zeta}+
\frac{1}{6a^2} \partial^i \epsilon^0_{(1)} \, \partial_i \epsilon^0_{(1)}
-\frac{1}{6a} \partial^i \beta \partial_i \epsilon^0_{(1)}
+\frac{1}{6} \nabla^2 \epsilon^s_{(2)}
\end{eqnarray}

We are now ready to compute the expansion rate $\tilde{\Theta}$, which
is a gauge dependent observable, in the unperturbed (uniform) field
gauge (UFG), while neglecting vector and tensor contributions and only keeping
terms up to second order:
\begin{eqnarray}
\tilde{\Theta}
&=& 3 \tilde H - 3 \tilde H \alphat +\frac{1}{a} \tilde{\nabla}^2 \betat - 3
{\tilde{\psi}}'
- 3 \tilde H 
\alphat^{(2)} + \frac{1}{a} \tilde{\nabla}^2 \betat^{(2)} - 
3 \tilde{\psi}^{(2)}{}'
\nonumber \\
& &
+\frac{9}{2} \tilde H \alphat^2 + 3 \alphat \tilde{\psi}' - 6 \tilde{\psi}
\tilde{\psi}' + 
\frac{1}{a}
 \vec{\tilde{\nabla}}\alphat \cdot \vec{\tilde{\nabla}} \betat 
- \frac{3}{a} \vec{\tilde{\nabla}}\betat \cdot \vec{\tilde{\nabla}} \tilde{\psi}
+ \frac{1}{4}\vec{\tilde{\nabla}}\betat \cdot \vec{\tilde{\nabla}} \betat' \, .
\label{tilde_theta}
\end{eqnarray}

Let us observe that the prime denotes the derivative with respect to
 $\tilde{t}$, and the relation between the two time derivatives may be written as
\begin{eqnarray}
\frac{\partial}{\partial \tilde{t}} &=& \left[1 - 
\frac{\dot{\zeta}}{H} -\epsilon \zeta+\left(-\frac{1}{2}\epsilon^2-\frac{1}{4}
\frac{\ddot{H}^2}{\dot{H}^2 H^2}+\frac{1}{4\dot{H} H^2}\frac{d^3 }{d t^3}(H)
\right)\zeta^2+
\left(\frac{\epsilon}{H}+\frac{1}{2}\frac{\ddot{H}}{\dot{H}H^2}\right)
\zeta\dot{\zeta}+\frac{1}{H^2}\dot{\zeta}^2\right]\frac{\partial}{\partial t}- \nonumber \\
  & & -\frac{1}{2} \partial_i \dot \epsilon^s_{(2)} \frac{\partial}{\partial x_i}  \,.
\label{time_rel}
\end{eqnarray}
Clearly the two time derivatives coincide on acting on second order terms
if one does not go beyond such an approximation, however
one must exercise care for terms of lower order on going from one
description to the other.

Further the difference beetwen the spatial derivatives is given by
\be
 \frac{\partial}{\partial \tilde{x}^i}= \frac{\partial}{\partial x^i}- \partial_i \epsilon^0_{(1)} 
\frac{\partial}{\partial t}
\ee
which, however, is not important since we shall neglect spatial derivatives.

One may wish to write the corrections to the
unperturbed $\tilde{\Theta}=3 \tilde{H}$ in the $\tilde{x}$ frame,
while expressing the corrections in terms of the already known
quantities (in the $x$ frame).
For this case one must be carefull in analyzing to second order the
fourth term of the above expression,
$-3 \partial \tilde{\psi}/ \partial \tilde{t}$.

To leading order, on averaging and neglecting terms with spatial
derivatives, we have:
\begin{eqnarray}
\langle \tilde{\Theta} \rangle &=&
3 \tilde H \left[ 1- \langle \alphat^{(2)} \rangle +\frac{3}{2}
\langle \alphat^2 \rangle +
\frac{1}{\tilde H} \langle \alphat \dot{\tilde{\psi}} \rangle-
\frac{1}{\tilde H} \langle \dot{\tilde{\psi}}^{(2)} \rangle-
\frac{2}{\tilde H} \langle \tilde{\psi} \dot{\tilde{\psi}} \rangle
-\frac{1}{\tilde H} \langle
\frac{\partial t}{\partial \tilde{t}} \dot{\tilde{\psi}} \rangle \right]
\nonumber \\
&\simeq& 3 \tilde H \left[1+\frac{\epsilon}{H} \langle \zeta \dot{\zeta}
  \rangle +\frac{1}{H^2} \langle \dot{\zeta}^2 \rangle
-\frac{1}{4} \frac{\ddot{H}}{H^3}  \langle \zeta^2 \rangle 
\right]
\nonumber \\
&=&  3 \tilde H \left[ 1 +\frac{1}{2 M_{pl}^2}
\left( -\frac{1}{H} \langle \varphi \dot{\varphi} \rangle
-\frac{1}{\dot{H}}\langle {\dot{\varphi}}^2
\rangle +{\cal O} ( \epsilon^2 \langle \varphi^2 \rangle)
\right)\right]
= 3 \tilde H \left[ 1 + \frac{1}{2 M_{pl}^2} {\cal O}
( \epsilon, \epsilon^2 \langle \varphi^2 \rangle) \right]\, .
\label{new_theta_nuovo_gauge}
\end{eqnarray}
where on going from the first to the second line the terms
${\cal O} ( \zeta \dot \zeta )$ cancel.
The infrared contributions to the terms in the last two lines vanish,
leading to corrections to the expansion rate in the UFG gauge of order
${\cal O} (\epsilon)$,
which are of the same order as the neglected gradient terms.
The corrections to the expansion rate in the two frames are therefore
of the same order in the slow-roll parameters.

If we choose to rewrite the averaged expression for $\tilde \Theta$ in
terms of $t$, starting from Eq. (\ref{tilde_theta})~\footnote{
Since the term $-3 \tilde{H} \tilde \alpha$ also has to be taken in to
account. Nonetheless one finds that such a term is, again, negligible and that
the leading result can be obtained starting directly from  Eq.
(\ref{new_theta_nuovo_gauge}).}
and using the relation in Eq. (\ref{time_rel})
(omitting negligible spatial derivatives) to also compute
$\tilde{H}(\tilde t)= H(t) \partial t /\partial {\tilde t}$, one finds 

\be
\langle \tilde{\Theta} \rangle = 3 H \left( 1 - \epsilon
\frac{\langle \varphi^2 \rangle}{4 M_{\rm pl}^2}
\right) \,,
\label{thetaresult_UFG}
\ee
which is completely consistent with our result in 
Eq. (\ref{thetaresult}) and with the fact that
$\Theta$ is a scalar (but not gauge invariant), which, therefore,
transforms according to:
\be
\tilde \Theta (\tilde x^\mu) = \Theta (x^\mu) \,.
\ee

We have then checked that the back-reaction of scalar cosmological
fluctuations  is ${\cal O} (\epsilon)$ in the expansion rate:
such a correction indeed vanishes
in the de Sitter limit, where scalar perturbations are absent. 
One may find a suitable gauge - for instance the UFG, which is valid
only during  slow-rollover -, in which the back-reaction does not grow in time 
(because of cancellations of the infrared pieces). However, such a gauge seems 
inconvenient since the inflaton cannot be used as a clock in the subsequent 
stage of coherent oscillations. 
We feel that it is more proper to ask what the time behavior of 
back-reaction is in a frame which can be regularly continued rather than the 
opposite.
Indeed, one of the advantages of the UCG is the possibility of continuing 
the calculation through the oscillatory phase of the inflaton 
\cite{reheating}, which is not easy to do in the UFG. It would be 
interesting to check if the same conclusions are obtained for the
non-local quantity used in \cite{nonlocal}.

\end{document}